\documentclass[10pt]{iopart}
\usepackage{graphicx}
\usepackage{epsfig}
\usepackage{epsf}
\begin{document}

\title[$\Lambda(1520)$ production at SPS and RHIC energies]{$\Lambda$(1520) production at SPS and RHIC Energies}
\author{Christina Markert \footnote[3]{E-mail: markert@star.physics.yale.edu} for the STAR collaboration
}

\address{Yale University, New Haven, Connecticut 06520;}
\address{supported by the Humboldt Foundation, Germany}

\begin{abstract}
The recent preliminary results from central Au+Au collisions at
$\sqrt{s_{\rm NN}} = $ 130 GeV from the STAR experiment at RHIC
are presented and discussed along with the results on the
$\Lambda$(1520) production in cental Pb+Pb and p+p collisions at
$\sqrt{s} = $ 17 GeV from the NA49 experiment at the SPS. The
$\Lambda$(1520) is measured with the invariant mass reconstruction
of the decay products in the hadronic channel (K$^{-}$, p). The
mean $\Lambda$(1520) multiplicity scaled by the number of
participants decreases from p+p to Pb+Pb collisions at the same
energy of $\sqrt{s}$~=~17 GeV. An upper limit estimate of the
multiplicity from the first measurement at $\sqrt{s_{\rm NN}} = $
130 GeV shows the same trend. Comparisons with model predictions
provides an indication of possible medium effects on the resonance
and their decay daughters.

\end{abstract}

\section{Introduction}

The $\Lambda$(1520) is a baryon resonance with a mass of m =
1519.5 MeV/c$^{2}$ and a width in vacuum of $\Gamma$ = 15.6 MeV.
This corresponds to a lifetime of $\tau$ = 13~fm/c which is of the
same order as the fireball source. The measurement of
$\Lambda$(1520) in heavy ion collisions may provide information
about the early state of the expanding source in terms of the
influence of the medium on the resonance and decay particles.
Comparisons between p+p and A+A interactions may show directly the
influence of the medium. However from these results it is not easy
to distinguish between a medium effect on the resonance or their
decay products. Additional multiplicity measurements from other
short lifetime resonances allow us to make an estimate of the
chemical freeze out temperature and lifetime of the fireball if
only rescattering of the decay products in the medium is
considered. The possible change of the width of the resonance
signal, associated with a change in lifetime, can provide an
additional effect on the multiplicity of the resonance measured
with the invariant mass reconstruction method.

\section{Analysis}

\begin{figure}[h!]
\begin{minipage}[b]{0.5\linewidth}
 \centering
\includegraphics[width=0.74\textwidth]{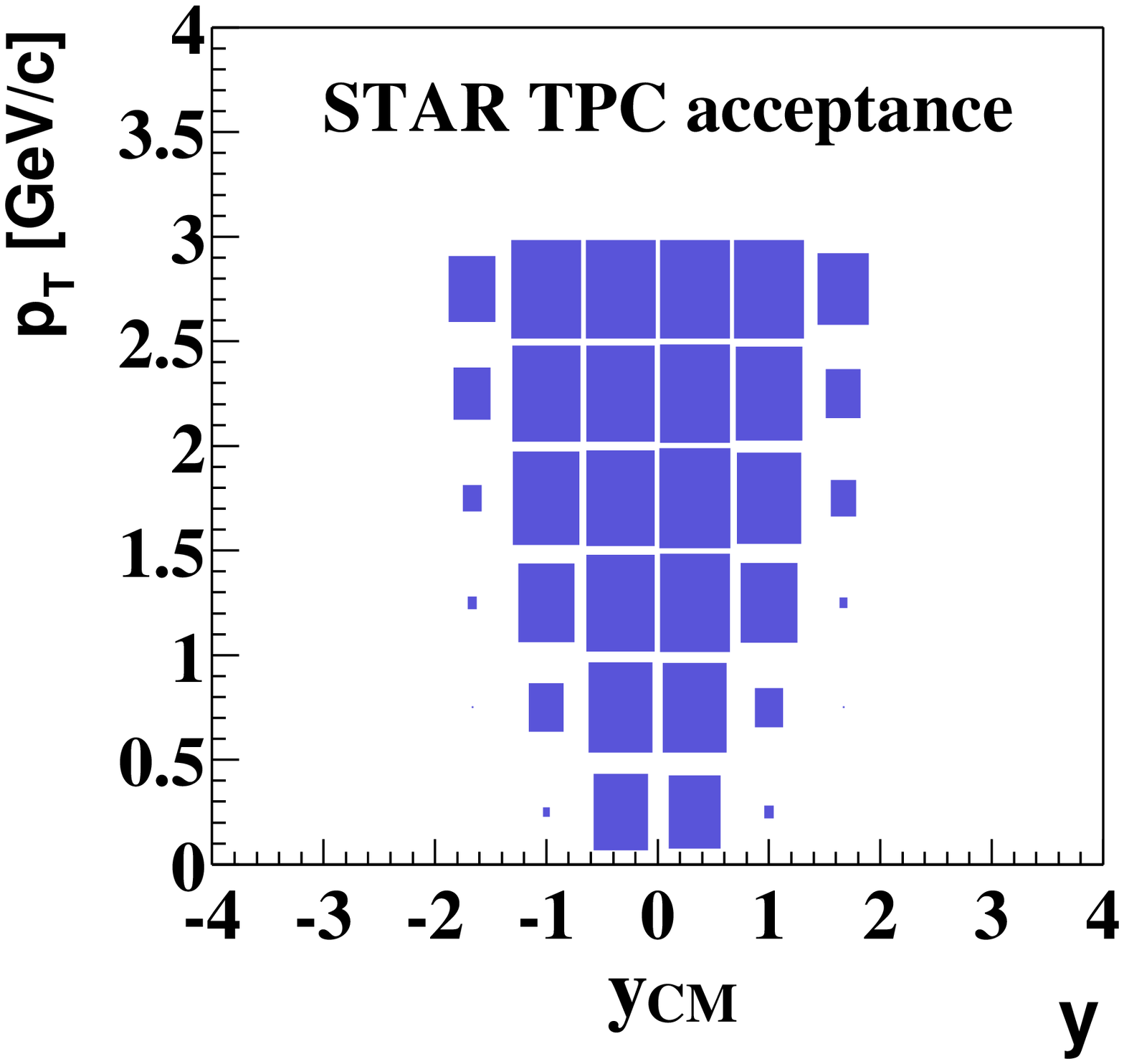}
 %\epsfbox{ratio_ypt_sqm2_2b.eps}
 %\caption{This is a caption.}
 \end{minipage}
 \hspace{-1cm}
 \begin{minipage}[b]{0.5\linewidth}
 \includegraphics[width=0.71\textwidth]{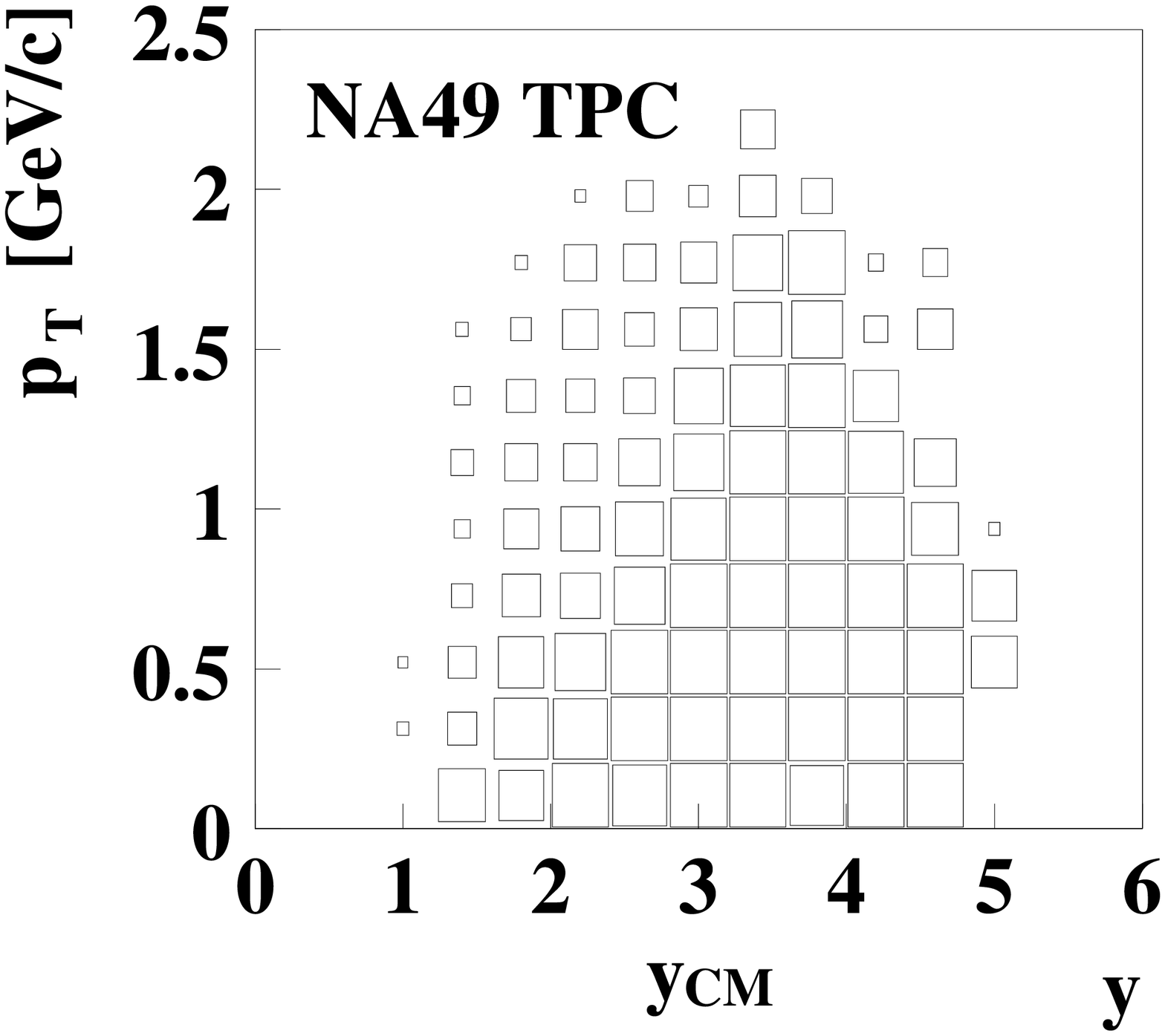}
 %\caption{This is a caption.}
\end{minipage}
\vspace{-0.2cm}
 \caption{Acceptance of the $\Lambda$(1520) in STAR
(left) and NA49 (right) TPCs in rapidity and transverse momentum.}
 \centering
 \label{acc}
\end{figure}

The STAR \cite{ack99} and the NA49 \cite{afa99} experimental setup
consists of large volume {\em Time Projection Chambers} (TPC)
which measure the momenta of charged particles and their energy
loss, $dE/dx$. A central trigger selects the 14\% (STAR) and 5\%
(NA49) most central inelastic interactions. The decay daughter
candidates for K$^{-}$ and p are selected via their momenta and
dE/dx. The $\Lambda$(1520) signal is obtained by the invariant
mass reconstruction of each pair combination and the subtraction
of a mixed event background estimated by combining candidates from
different events. The largest uncertainty of this analysis is
calculating the correction factor for a total (4$\pi$)
multiplicity in A+A collisions, where the phase space distribution
is not measured. The inverse slope parameters T are taken from the
measured $\Lambda$s, T = 350 MeV (STAR \cite{mat01}) and T = 300
MeV (NA49 \cite{mis01}). The assumption for the rapidity
distribution is a gaussian with a width $\sigma$ = 1.5 $\pm$ 0.5
and $\sigma$ = 1.0 $\pm$ 0.5 for STAR and NA49 respectively. The
total mean acceptances in the TPCs are 0.51 $\pm$ 0.08 (STAR) and
0.82 $\pm$ 0.5 (NA49). Figure~\ref{acc} shows the acceptance
distributions from a simulation in phase space coordinates y and
p$_{T}$ for the STAR and NA49 TPCs.

\vspace{-0.2cm}

\section*{p+p at $\sqrt{s} = $ 17 GeV}

\begin{figure}[h!]
\begin{minipage}[b]{0.5\linewidth}
 \centering
 \includegraphics[width=0.9\textwidth]{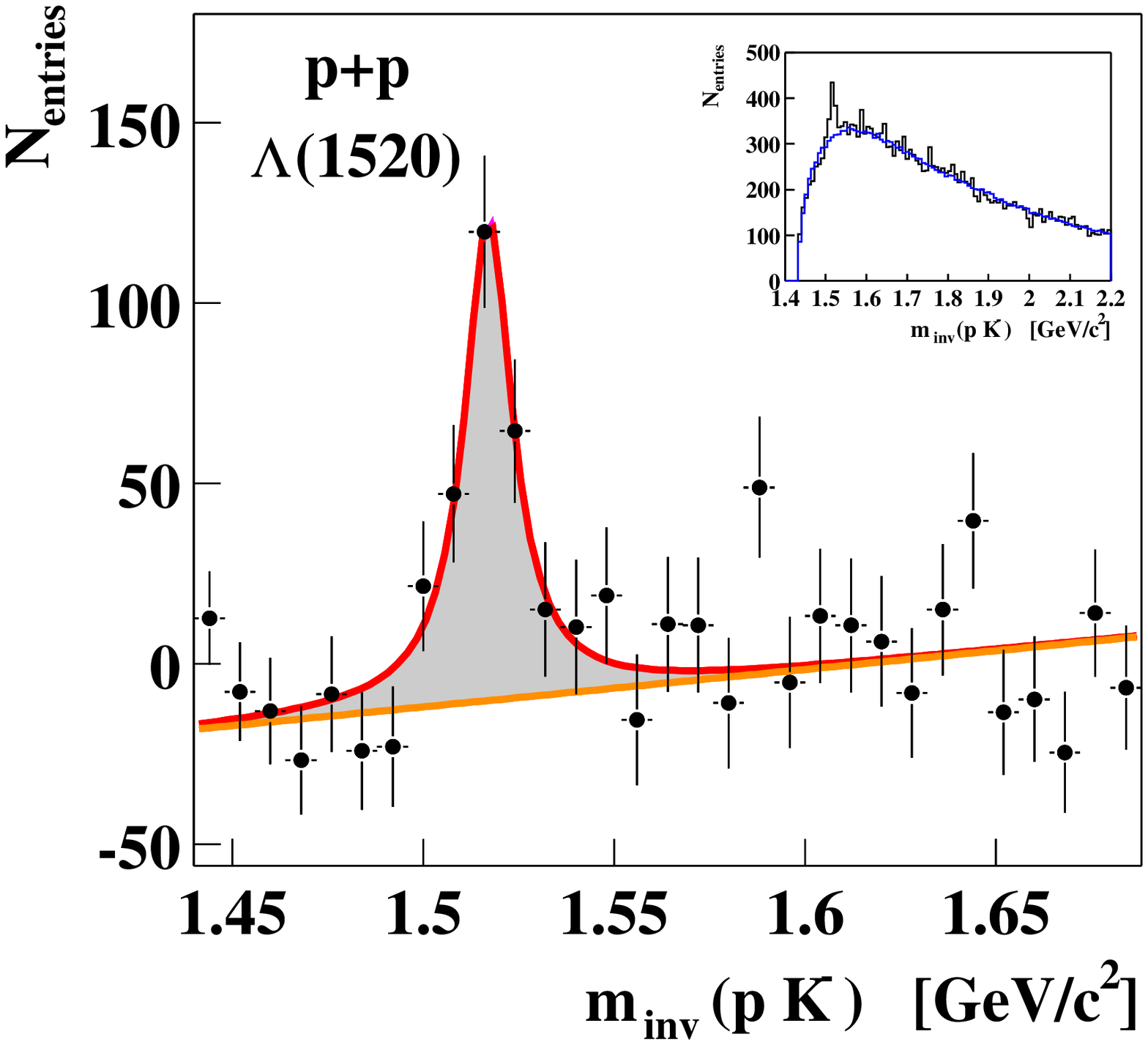}
 %\caption{$\Lambda$(1520) mass plot after mixed event background subtraction with a Breit-Wigner fit , upper plot: Mass plot before mixed event background subtraction.}
 \label{pp}
 \end{minipage}
 \hspace{-0.5cm}
 \begin{minipage}[b]{0.5\linewidth}
 \centering
 \includegraphics[width=0.9\textwidth]{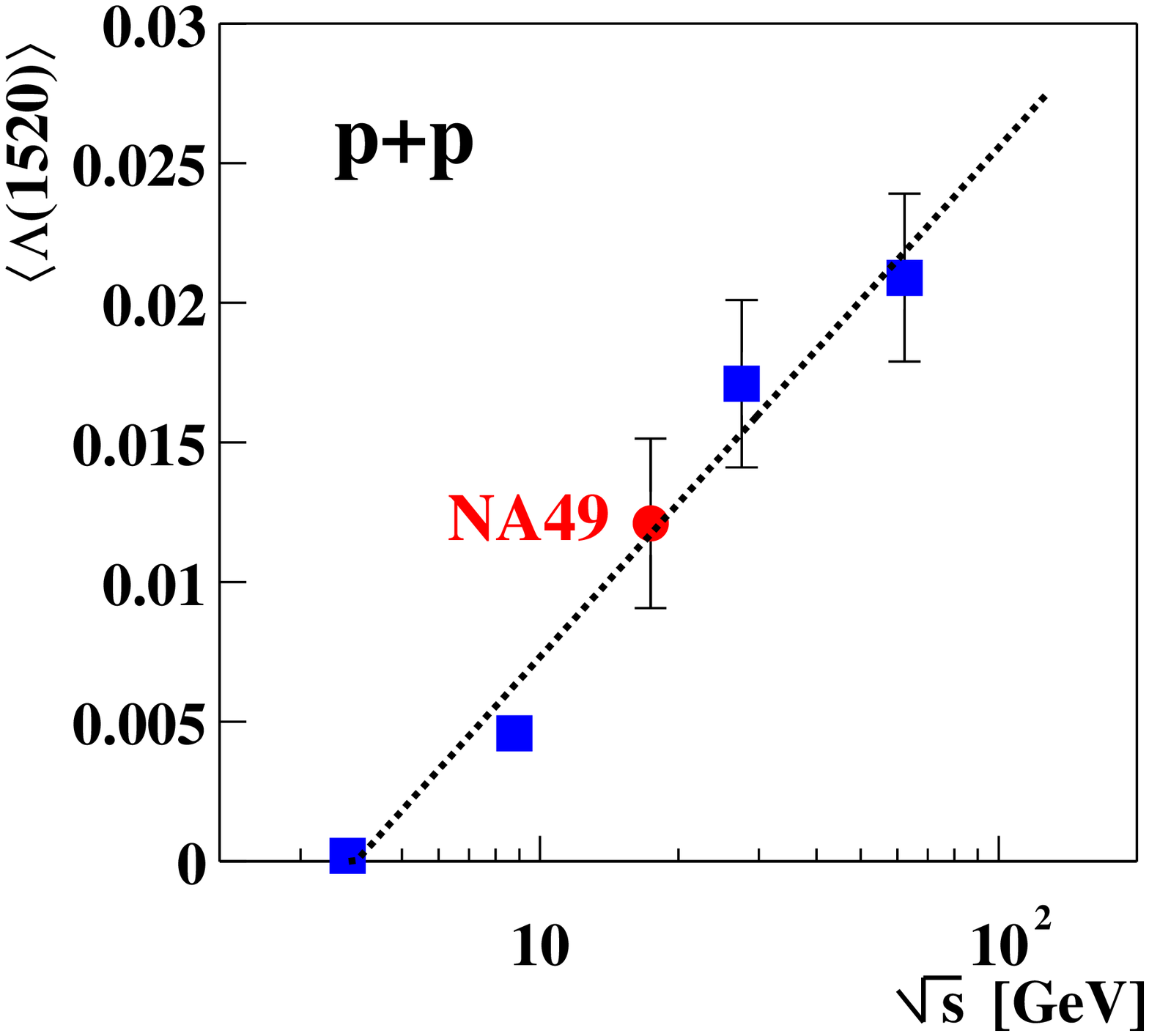}
 %\caption{$\Lambda$(1520) multiplicity in elementary p+p collisions for different energies.}
 \label{ppenergie}
\end{minipage}
\vspace{-0.2cm}
 \caption{Left: $\Lambda$(1520) mass plot after mixed event
background subtraction with a Breit-Wigner function. Insert plot:
Mass plot before mixed event background subtraction. Right:
$\Lambda$(1520) multiplicity in elementary p+p collisions for
various energies \cite{ans74,kra88,agu91,bob83}.}
\end{figure}

The invariant mass distribution from 400,000 p+p collisions is
shown in Figure~\ref{pp} (left) \cite{fri01,mar01}. The
Breit-Wigner fit parameters mass m = 1517.1 $\pm$ 1.5 MeV/c$^{2}$
and width $\Gamma$ = 15.4 $\pm$ 3.8 MeV/c$^{2}$ are consistent
within errors with the values from the Particle Data Group
\cite{pdg98}. The mean multiplicity is
$\langle\Lambda$(1520)$\rangle$ = 0.0121 $\pm$ 0.003. A comparison
with measurements at different energies in Figure~\ref{pp}(right)
shows good agreement with the energy dependence of the particle
production.

\vspace{-0.2cm}

\section*{Pb+Pb at $\sqrt{s} = $ 17 GeV}

\begin{figure}[h!]
 \centering
 \includegraphics[width=0.55\textwidth]{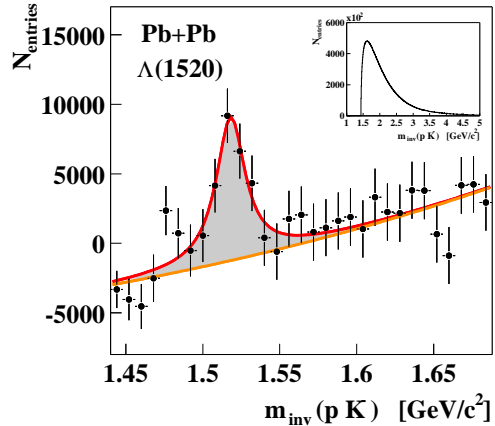}
 %\epsfbox{lam_pbpb.eps,height=10mm}
 %\epsfig{file=lam_pbpb.eps,height=100mm}
 \caption{$\Lambda$(1520) mass plot after mixed event
background subtraction with a Breit-Wigner fit , insert plot: Mass
plot before mixed event background subtraction from NA49 at
$\sqrt{s} = $ 17 GeV.}
 \label{pbpb}
\end{figure}

Figure~\ref{pbpb} shows the invariant mass distribution with
400,000 events from Pb+Pb collisions \cite{fri01,mar01}. The
Breit-Wigner fit parameters for the mass m = 1518.1 $\pm$ 2.0
MeV/c$^{2}$ and width $\Gamma$ = 22.7 $\pm$ 6.5 MeV/c$^{2}$. The
$\Gamma$ seems a little broad but is consistent within errors with
the values from the Particle Data Group \cite{pdg98}. This leads
to a mean multiplicity $\langle\Lambda$(1520)$\rangle$ = 1.45
$~\pm$~0.29~$\pm$~0.28 with statistical and systematical errors.

\section{Discussion of $\Lambda$(1520) production at $\sqrt{s}
= $ 17 GeV} If we scale the $\Lambda$(1520) multiplicity from p+p
collisions with the number of participants for central Pb+Pb
collisions (360/2) we get a prediction of
$\langle\Lambda$(1520)$\rangle$~=~2.2. If we then scale with the
known $\Lambda$ strangeness enhancement factor ($\sim$ 2-2.5)
\cite{gaz99} the yield becomes $\langle\Lambda$(1520)$\rangle$
$\approx$ 5. This number is close to the predictions from thermal
model calculations with $\langle\Lambda$(1520)$\rangle$ = 3.5
($\gamma_{s}$ = 0.7) \cite{bec98} and
$\langle\Lambda$(1520)$\rangle$ = 5.2 ($\gamma_{s}$ = 1)
\cite{pbm99}. The measured yield in central Pb+Pb collisions is
$\langle\Lambda$(1520)$\rangle$~=1.45~$\pm$~0.40 which is a factor
of 2-3 lower than the expected value. This trend of signal
suppression is also visible in the $\Lambda$(1520)/$\Lambda$ and
$\Lambda$(1520)/$\pi$ particle ratios.

\begin{figure}
\begin{minipage}[b]{0.5\linewidth}
 \centering
 \includegraphics[width=0.85\textwidth]{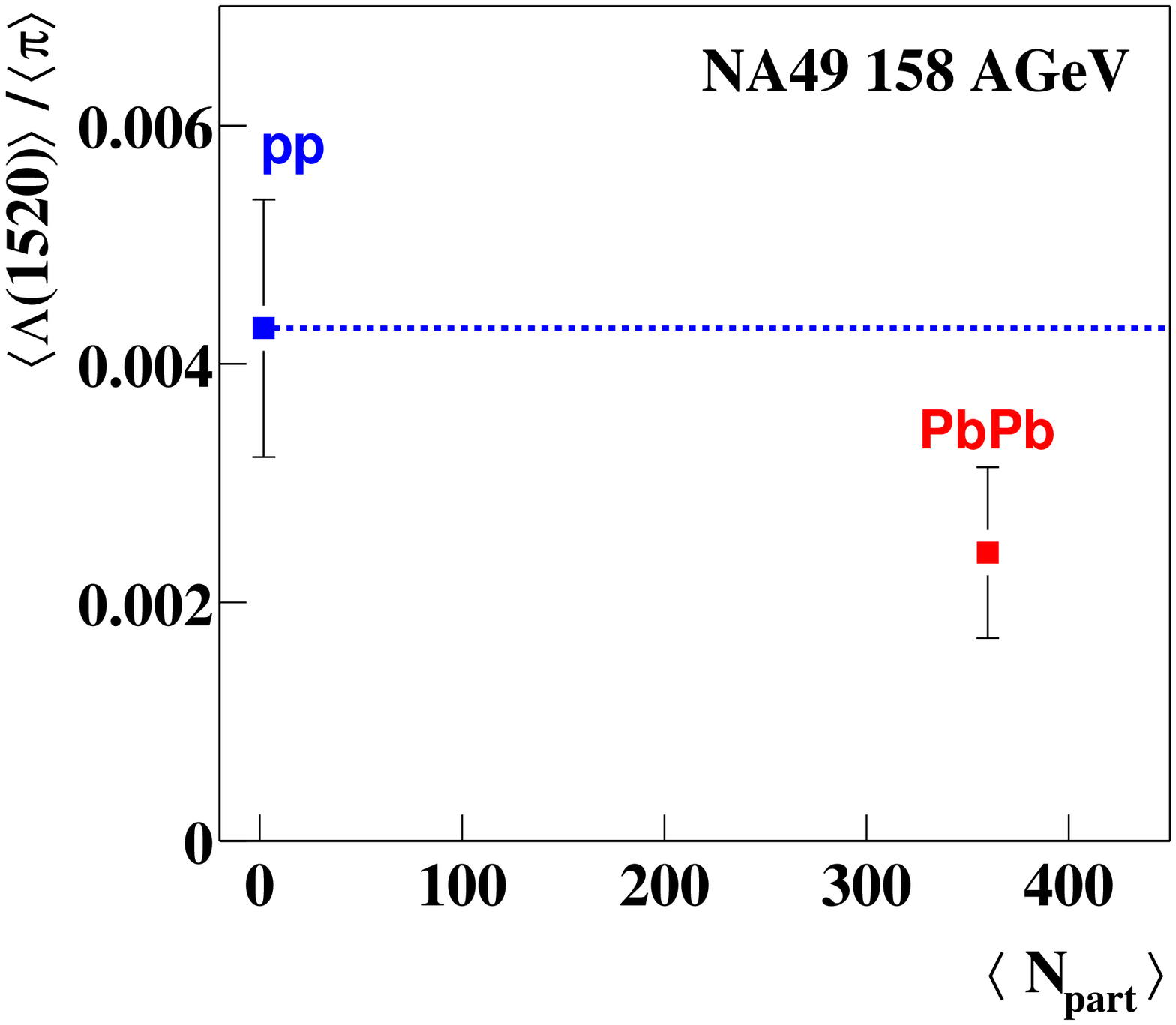}
 %\caption{This is a caption.}
 \label{lampion}
 \end{minipage}
 \hspace{-0.5cm}
 \begin{minipage}[b]{0.5\linewidth}
 \centering
 \includegraphics[width=0.85\textwidth]{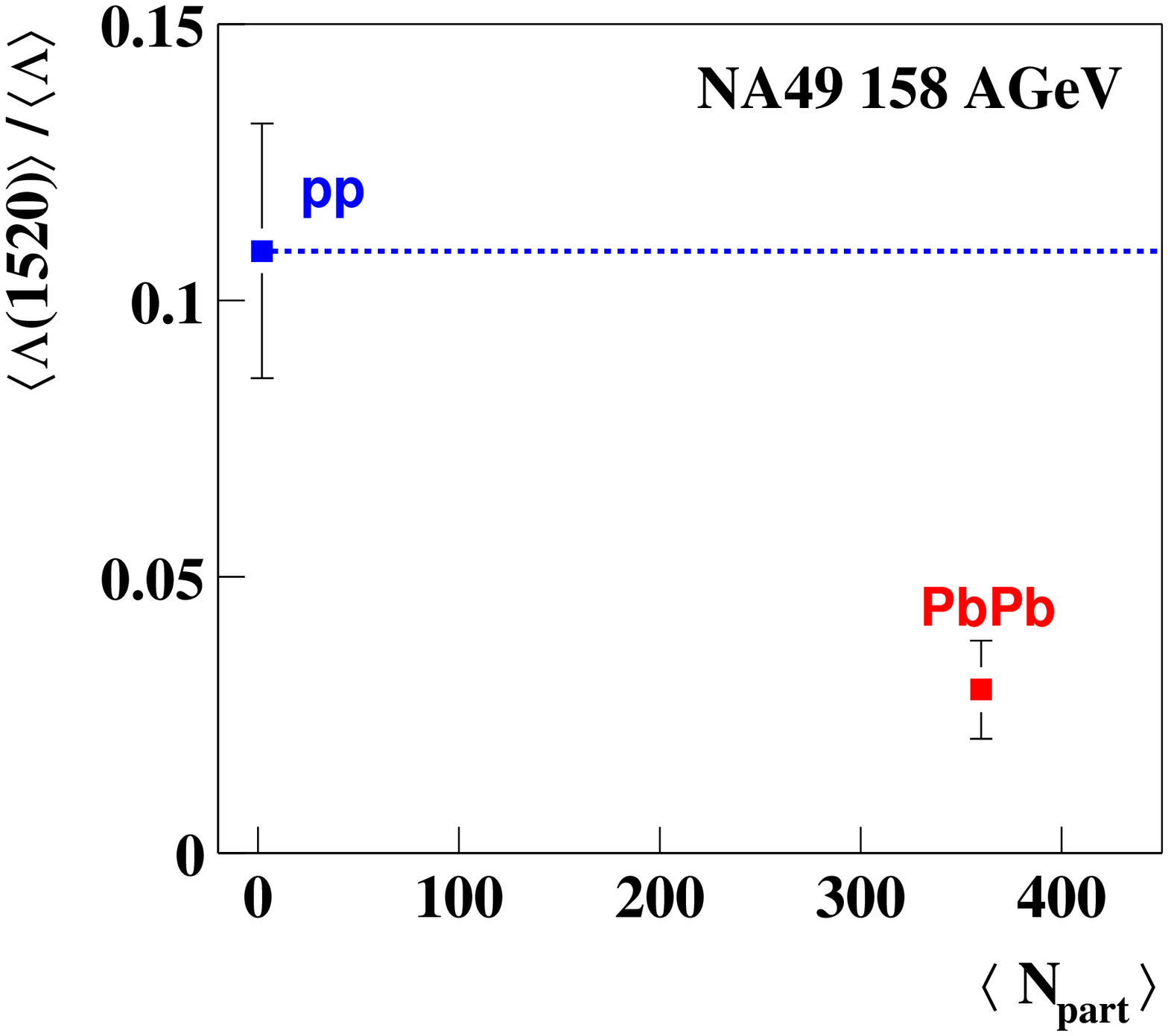}
 %\caption{This is a caption.}
 \label{lamlam}
\end{minipage}
\caption{Left: $\langle\Lambda$(1520)$\rangle$/$\pi$ ratio as a
function of number of participants. Right:
$\langle\Lambda$(1520)$\rangle$/$\Lambda$ ratio as a function of
number of participants. Data points are p+p and central Pb+Pb
collisions from NA49.}
\label{ratios}
\end{figure}

\begin{figure}[h!]
 \centering
 \includegraphics[width=0.45\textwidth]{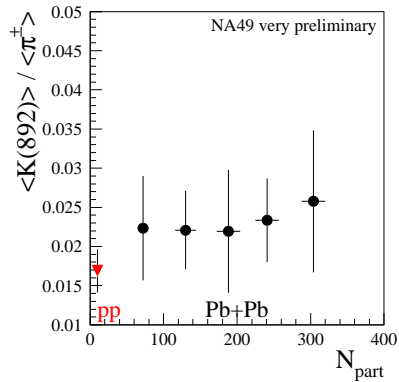}
 \caption{K$^{*}$(892)/$\pi$ as a function of participants for p+p and Pb+Pb collisions at 158~AGeV.}
 \label{kstar}
\end{figure}

The particle ratios
$\langle\Lambda$(1520)$\rangle$/$\langle\pi\rangle$ in
Figure~\ref{ratios} (left) and
$\langle\Lambda$(1520)$\rangle$/$\langle\Lambda\rangle$ in
Figure~\ref{ratios} (right) decrease from p+p to Pb+Pb collisions.
If we assume that the
$\langle\Lambda$(1520)$\rangle$/$\langle\Lambda\rangle$ ratio were
produced thermally the chemical freeze out temperature would be
smaller than 100 MeV from the model predictions. This is in
contradiction to the freeze out temperature calculated from other
particles \cite{bec98}. F. Becattini pointed out in his talk at
this conference that the $\Lambda$(1520) multiplicity at SPS
energies cannot be described by his thermal model. The question is
what can cause this signal loss during the expansion of the
fireball source. There are several ideas of possible medium
effects which can cause a reduction of the measured
$\Lambda$(1520) resonance. One possible medium effect on the decay
products is their rescattering, which changes their momentum and
energy. The UrQMD model calculation gives a signal loss of 50\%
for the $\Lambda$(1520) \cite{sof01} as a result of rescattering
of the K$^{-}$ and the protons. If only the rescattering effect is
present, it should also be seen for the K$^{*}$(892), whose
lifetime is 3 times shorter. However from NA49 the
K$^{*}$(892)/$\pi$ measurement \cite{fri01} shows no visible
suppression within errors (see Figure~\ref{kstar}).

\begin{figure}[h!]
 \centering
 \includegraphics[width=0.55\textwidth,angle=0]{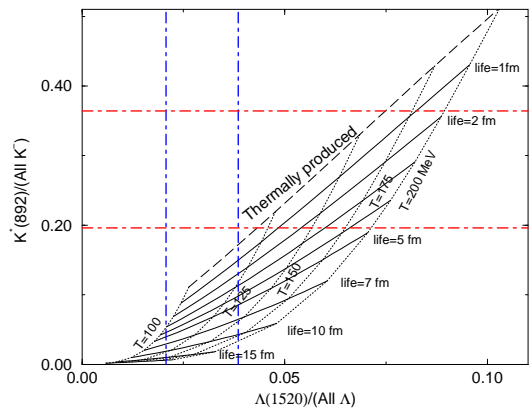}
 \caption{Dependence of lifetime and temperature of a fireball source given by two particle ratios $\Lambda$(1520)/$\Lambda$ and
K$^{*}$(892)/K \cite{tor01,tor01a}.}
 \label{raf}
\end{figure}

Model calculations shown in Figure~\ref{raf} \cite{tor01,tor01a}
including the lifetime of the fireball source and the chemical
freeze out temperature by comparing two particle ratios namely
$\Lambda$(1520)/$\Lambda$ and K$^{*}$(892)/K shows that there is
only agreement between the two ratios if the temperature is lower
than 100 MeV and the time between the chemical and thermal freeze
out is close to zero. The low temperature is in contradiction with
the thermal freeze out temperature, which is around 170 MeV
\cite{bec98}. A conclusion from this model is that the
rescattering of the decay particles in medium may not be the only
effect.

\newpage

Another possibility which would result in a reduction of the
measured $\Lambda$(1520) signal can be a medium effect on the
resonance itself. A broader width of the resonance in medium
corresponds to a shorter lifetime and this in terms causes a
larger signal loss due to rescattering of the decay products. A
prediction based on relativistic chiral SU(3) dynamic calculations
\cite{lut01,lut2} gives a 100 MeV broadening of the
$\Lambda$(1520) resonance and a mass shift of about 100 MeV to
lower masses in medium at a density of $\rho$ = 0.17~fm$^{-3}$.
The fact that the $\Lambda$(1520) signal from Pb+Pb reactions from
NA49 shows no width broadening within the errors, indicates that
only the decay daughters which come from a $\Lambda$(1520)
resonance with normal width survive and decay at the end of the
fireball evolution. In this conference a talk from E.E.
Kolomeitsev shows that this medium effect is not only able to
describe the signal loss it may also describe a possible change in
the slopes of the p$_{\rm T}$ distribution. This was shown for the
$\phi$ particle in the hadronic and the leptonic channel. The
calculations are only done for the $\phi$. The $\Lambda$(1520) and
the K$^{*}$(892) are in progress. To gain some confidence in the
signal suppression of the $\Lambda$(1520) resonance due to
in-medium effects it is necessary to measure additional resonances
(e.g. $\Sigma(1385)$) and take additional measurements at
different energies.

\section{Au+Au at $\sqrt{s_{\rm NN}} = $ 130 GeV }
\begin{figure}[h!]
\begin{minipage}[b]{0.5\linewidth}
 \centering
 \includegraphics[width=0.9\textwidth]{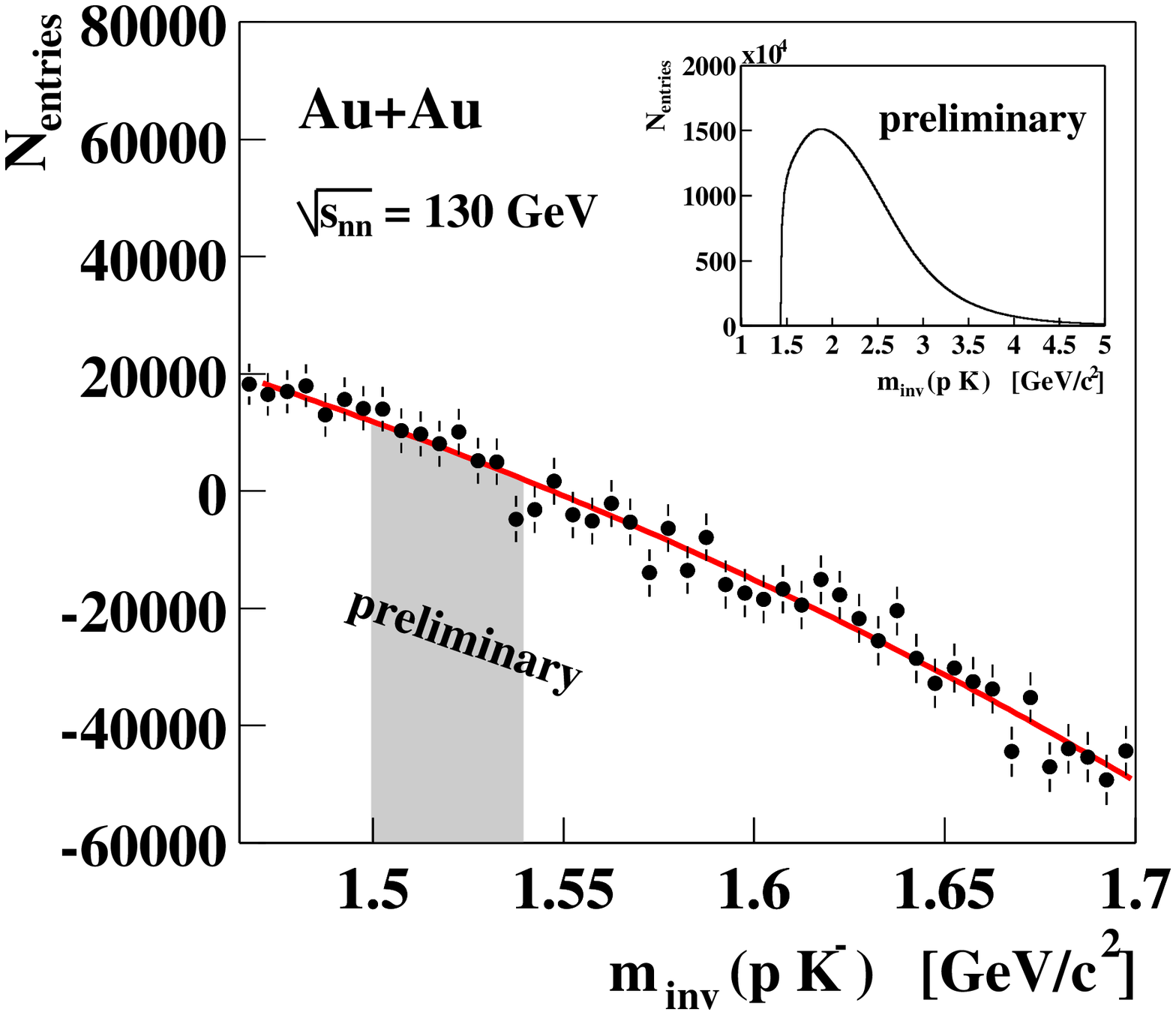}
 %\caption{This is a caption.}
 \end{minipage}
 \hspace{-0.5cm}
 \begin{minipage}[b]{0.5\linewidth}
 \centering
 \includegraphics[width=0.9\textwidth]{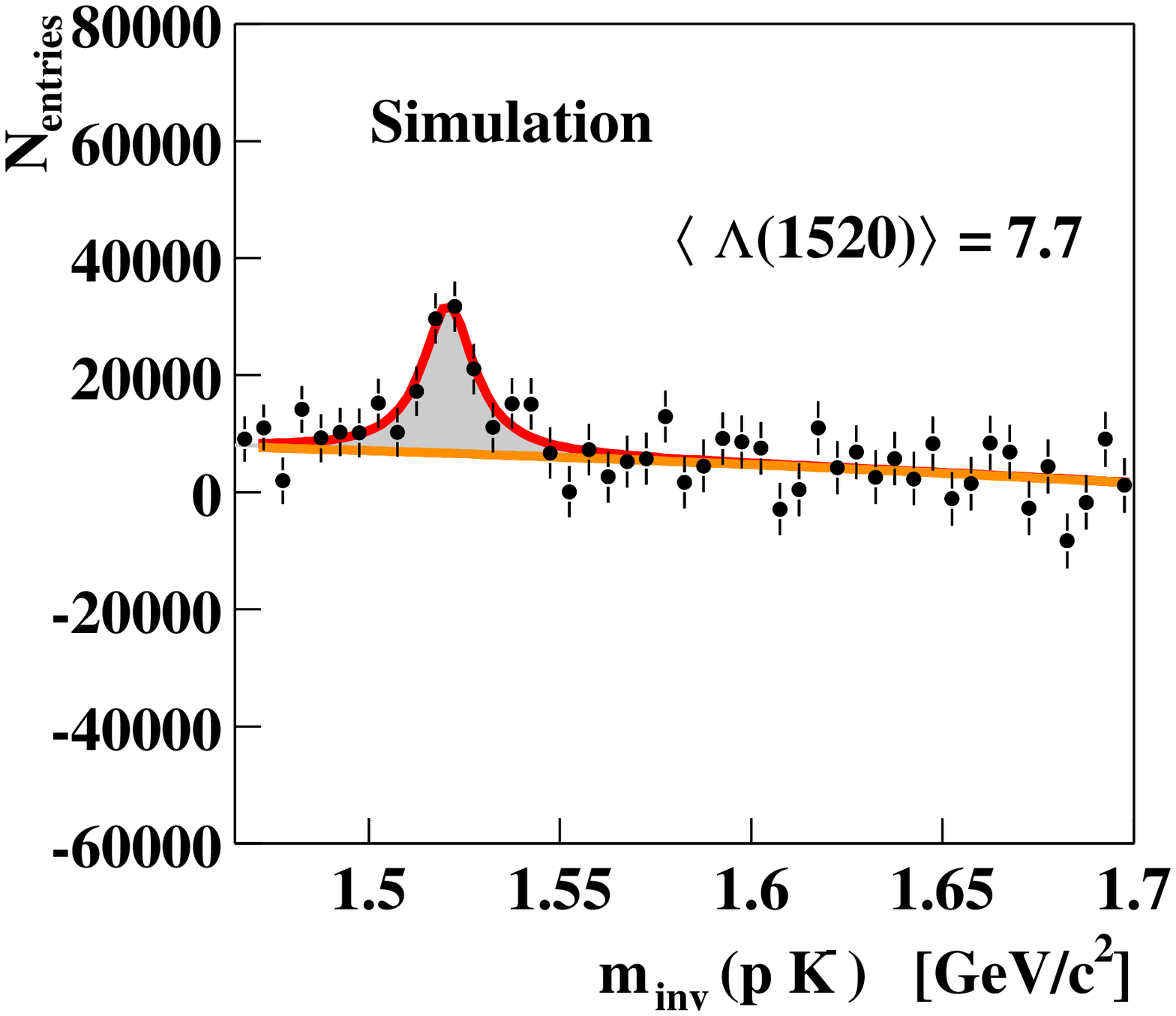}
 %\caption{This is a caption.}
\end{minipage}
\caption{Left: Invariant mass plot for K$^{-}$ and p pairs after
mixed event background subtraction. Insert plot: Mass plot before
mixed event background subtraction. Right: Simulation for
$\langle\Lambda$(1520)$\rangle$ = 7.7.} \label{starauau}
\end{figure}

The RHIC data from 2000 are used to look for the $\Lambda$(1520)
at higher energy. The analysis is done with 370,000 central Au+Au
events. Figure~\ref{starauau} (left) shows the invariant mass
distribution of the selected decay particles (insert plot) and the
mixed event background subtracted histogram. There is no clear
$\Lambda$(1520) signal visible, also the background from the mixed
event method is imperfect. The counts in the invariant mass region
where we expect the signal multiplied by the correction factor
give a number which is consistent with zero within statistical and
systematical errors. The preliminary total mean multiplicity is
$\langle\Lambda$(1520)$\rangle$ = 0.92 $\pm$ 0.81 $\pm$ 0.83. From
the statistical errors we can estimate how sensitive we are to
$\Lambda$(1520) production and exclude a multiplicity using the
upper limit method. Here $\langle\Lambda$(1520)$\rangle$ $<$ 4.2
at the 95\% confidence level. The expected multiplicity from
extrapolated elementary p+p reactions including an addition factor
of 2 for strangeness enhancement is $\sim$ 7.7. The upper limit
estimate suggests that at RHIC energies we see the same trend of
signal loss as at SPS energies. The simulation in
Figure~\ref{starauau} (right) shows the predicted invariant mass
signal from a statistical analysis if the
$\langle\Lambda$(1520)$\rangle$ yield were 7.7.

\section{Conclusions}

The $\Lambda$(1520) multiplicity measured from the NA49 experiment
at the SPS in p+p collisions at $\sqrt{s} = $ 17 GeV is in good
agreement with the data from the literature. The measured
multiplicity in Pb+Pb at $\sqrt{s} = $~17 GeV is
$\langle\Lambda$(1520)$\rangle$ = 1.45 $\pm$ 0.40 which cannot de
described by a thermal model only. A suppression factor of 2 is
needed to reproduce the measured multiplicity. A comparison with
the K$^{*}$(892) resonance gives indications of medium effects on
the $\Lambda$(1520) resonance itself. The first measurement from
the STAR experiment at RHIC energies $\sqrt{s_{\rm NN}} = $ 130
GeV gives an upper limit of $\langle\Lambda$(1520)$\rangle$ $<$
4.2, which follows the same suppression trend as the SPS results.
The question why the $\Lambda$(1520) multiplicity is lower than
expected is unanswered yet, but many interesting theoretical ideas
have been proposed. We hope to extract a signal from STAR with the
data from the year 2001 where are about 3.5 million central Au+Au
events at $\sqrt{s_{\rm NN}} = $ 200 GeV have been recorded.

\section*{Acknowledgments} We wish to thank the RHIC Operations
Group and the RHIC Computing Facility at Brookhaven National
Laboratory, and the National Energy Research Scientific Computing
Center at Lawrence Berkeley National Laboratory for their support.
This work was supported by the Division of Nuclear Physics and the
Division of High Energy Physics of the Office of Science of the
U.S.Department of Energy, the United States National Science
Foundation, the Bundesministerium fuer Bildung und Forschung of
Germany, the Institut National de la Physique Nucleaire et de la
Physique des Particules of France, the United Kingdom Engineering
and Physical Sciences Research Council, Fundacao de Amparo a
Pesquisa do Estado de Sao Paulo, Brazil, the Russian Ministry of
Science and Technology and the Ministry of Education of China and
the National Science Foundation of China.

\end{document}